\documentclass[10pt]{IEEEtran}
\usepackage{url}
\usepackage[utf8]{inputenc}
\usepackage{xcolor}
\usepackage{amsmath}
\usepackage{amssymb}

\usepackage[acronyms,nonumberlist,nopostdot,nomain,nogroupskip]{glossaries}
\usepackage{tablefootnote}
\usepackage{booktabs}

\usepackage{tabularx}

\usepackage{tikz}
\usepackage{pgfplots}
\pgfplotsset{compat=newest}
\pgfplotsset{plot coordinates/math parser=false}
\newlength\fheight
\newlength\fwidth
\usetikzlibrary{plotmarks,patterns,decorations.pathreplacing,backgrounds,calc,arrows,arrows.meta,spy,matrix}
\usepgfplotslibrary{patchplots,groupplots}
\usepackage{tikzscale}
\usepackage{hyperref}

\newif\ifexttikz
\exttikzfalse

\ifexttikz
	\usetikzlibrary{external}
\fi

\usepackage{multirow}
\usepackage{tkz-kiviat}

\usepackage[font=footnotesize]{subcaption}
\usepackage[font=footnotesize]{caption}

\usepackage{mathtools}

\newacronym{3gpp}{3GPP}{3rd Generation Partnership Project}
\newacronym{adc}{ADC}{Analog to Digital Converter}
\newacronym{5g}{5G}{5th generation}
\newacronym{aimd}{AIMD}{Additive Increase Multiplicative Decrease}
\newacronym{am}{AM}{Acknowledged Mode}
\newacronym{amc}{AMC}{Adaptive Modulation and Coding}
\newacronym{aqm}{AQM}{Active Queue Management}
\newacronym{awgn}{AGWN}{Additive White Gaussian Noise}
\newacronym{balia}{BALIA}{Balanced Link Adaptation}
\newacronym{bdp}{BDP}{Bandwidth-Delay Product}
\newacronym{bf}{BF}{Beamforming}
\newacronym{cc}{CC}{Congestion Control}
\newacronym{cdf}{CDF}{Cumulative Distribution Function}
\newacronym{cn}{CN}{Core Network}
\newacronym{cqi}{CQI}{Channel Quality Information}
\newacronym{cp}{CP}{Control Plane}
\newacronym{csirs}{CSI-RS}{Channel State Information - Reference Signal}
\newacronym{dc}{DC}{Dual Connectivity}
\newacronym{dce}{DCE}{Direct Code Execution}
\newacronym{dci}{DCI}{Downlink Control Information}
\newacronym{dl}{DL}{Downlink}
\newacronym{dmr}{DMR}{Deadline Miss Ratio}
\newacronym{dmrs}{DMRS}{DeModulation Reference Signal}
\newacronym{e2e}{E2E}{End-to-End}
\newacronym{ecn}{ECN}{Explicit Congestion Notification}
\newacronym{edf}{EDF}{Earliest Deadline First}
\newacronym{enb}{eNB}{evolved Node Base}
\newacronym{epc}{EPC}{Evolved Packet Core}
\newacronym{es}{ES}{Edge Server}
\newacronym{fdma}{FDMA}{Frequency Division Multiple Access}
\newacronym{fdd}{FDD}{Frequency Division Duplexing}
\newacronym[firstplural=Radio Access Technologies (RATs)]{rat}{RAT}{Radio Access Technology}
\newacronym{fs}{FS}{Fast Switching}
\newacronym{ftp}{FTP}{File Transfer Protocol}
\newacronym{gnb}{gNBs, the NR term for a base station}{Next Generation Node Bases}
\newacronym{harq}{HARQ}{Hybrid Automatic Repeat reQuest}
\newacronym{hetnet}{HetNet}{Heterogeneous Network}
\newacronym{hh}{HH}{Hard Handover}
\newacronym{hol}{HOL}{Head-of-Line}
\newacronym{ia}{IA}{Initial Access}
\newacronym{imt}{IMT}{International Mobile Telecommunication}
\newacronym{iot}{IoT}{Internet of Things}
\newacronym{los}{LOS}{Line-of-Sight}
\newacronym{lte}{LTE}{Long Term Evolution}
\newacronym{m2m}{M2M}{Machine to Machine}
\newacronym{mac}{MAC}{Medium Access Control}
\newacronym{mc}{MC}{Multi-Connectivity}
\newacronym{mcs}{MCS}{Modulation and Coding Scheme}
\newacronym{mec}{MEC}{Mobile Edge Cloud}
\newacronym{mi}{MI}{Mutual Information}
\newacronym{mimo}{MIMO}{Multiple Input, Multiple Output}
\newacronym{mmwave}{mmWave}{millimeter wave}
\newacronym{mptcp}{MPTCP}{Multipath TCP}
\newacronym{mr}{MR}{Maximum Rate}
\newacronym{mss}{MSS}{Maximum Segment Size}
\newacronym{mtd}{MTD}{Machine-Type Device}
\newacronym{mtu}{MTU}{Maximum Transmission Unit}
\newacronym{nfv}{NFV}{Network Function Virtualization}
\newacronym{nlos}{NLOS}{Non-Line-of-Sight}
\newacronym{nr}{NR}{New Radio}
\newacronym{ofdm}{OFDM}{Orthogonal Frequency Division Multiplexing}
\newacronym{pdcch}{PDCCH}{Physical Downlonk Control Channel}
\newacronym{pdcp}{PDCP}{Packet Data Convergence Protocol}
\newacronym{pdsch}{PDSCH}{Physical Downlink Shared Channel}
\newacronym{pdu}{PDU}{Packet Data Unit}
\newacronym{pf}{PF}{Proportional Fair}
\newacronym{pgw}{PGW}{Packet Gateway}
\newacronym{phy}{PHY}{Physical}
\newacronym{pbch}{PBCH}{Physical Broadcast Channel}
\newacronym[plural=\gls{mme}s,firstplural=Mobility Management Entities (MMEs)]{mme}{MME}{Mobility Management Entity}
\newacronym{prb}{PRB}{Physical Resource Block}
\newacronym{pss}{PSS}{Primary Synchronization Signal}
\newacronym{pucch}{PUCCH}{Physical Uplink Control Channel}
\newacronym{pusch}{PUSCH}{Physical Uplink Shared Channel}
\newacronym{rach}{RACH}{Random Access Channel}
\newacronym{ran}{RAN}{Radio Access Network}
\newacronym{red}{RED}{Random Early Detection}
\newacronym{rf}{RF}{Radio Frequency}
\newacronym{rlc}{RLC}{Radio Link Control}
\newacronym{rlf}{RLF}{Radio Link Failure}
\newacronym{rrc}{RRC}{Radio Resource Control}
\newacronym{rrm}{RRM}{Radio Resource Management}
\newacronym{rr}{RR}{Round Robin}
\newacronym{rs}{RS}{Remote Server}
\newacronym{rsrp}{RSRP}{Reference Signal Received Power}
\newacronym{rss}{RSS}{Received Signal Strength}
\newacronym{rtt}{RTT}{Round Trip Time}
\newacronym{rw}{RW}{Receive Window}
\newacronym{rx}{RX}{Receiver}
\newacronym{sa}{SA}{standalone}
\newacronym{sack}{SACK}{Selective Acknowledgment}
\newacronym{sap}{SAP}{Service Access Point}
\newacronym{sch}{SCH}{Secondary Cell Handover}
\newacronym{scoot}{SCOOT}{Split Cycle Offset Optimization Technique}
\newacronym{sdma}{SDMA}{Spatial Division Multiple Access}
\newacronym{sinr}{SINR}{Signal to Interference plus Noise Ratio}
\newacronym{sm}{SM}{Saturation Mode}
\newacronym{snr}{SNR}{Signal-to-Noise-Ratio}
\newacronym{son}{SON}{Self-Organizing Network}
\newacronym{ss}{SS}{Synchronization Signal}
\newacronym{srs}{SRS}{Sounding Reference Signal}
\newacronym{sss}{SSS}{Secondary Synchronization Signal}
\newacronym{tb}{TB}{Transport Block}
\newacronym{tcp}{TCP}{Transmission Control Protocol}
\newacronym{tdd}{TDD}{Time Division Duplexing}
\newacronym{tdma}{TDMA}{Time Division Multiple Access}
\newacronym{tfl}{TfL}{Transport for London}
\newacronym{tm}{TM}{Transparent Mode}
\newacronym{trp}{TRP}{Transmitter Receiver Pair}
\newacronym{tti}{TTI}{Transmission Time Interval}
\newacronym{ttt}{TTT}{Time-to-Trigger}
\newacronym{tx}{TX}{Transmitter}
\newacronym{ue}{UE}{User Equipment}
\newacronym{ul}{UL}{Uplink}
\newacronym{uml}{UML}{Unified Modeling Language}
\newacronym{um}{UM}{Unacknowledged Mode}
\newacronym{utc}{UTC}{Urban Traffic Control}
\newacronym{vm}{VM}{Virtual Machine}
\newacronym{rsrq}{RSRQ}{Reference Signal Received Quality}
\newacronym{rssi}{RSSI}{Received Signal Strength Indicator}
\newacronym{crs}{CRS}{Cell Reference Signal}
\newacronym{nsa}{NSA}{Non Stand Alone}
\newacronym{mrdc}{MR-DC}{Multi \gls{rat} \gls{dc}}
\newacronym{endc}{EN-DC}{E-UTRAN-\gls{nr} \gls{dc}}
\newacronym{5gc}{5GC}{5G Core}
\newacronym{si}{SI}{Study Item}
\newacronym{iab}{IAB}{Integrated Access and Backhaul}
\newacronym{wf}{WF}{Wired-first}
\newacronym{hqf}{HQF}{Highest-quality-first}
\newacronym{pa}{PA}{Position-aware}
\newacronym{mlr}{MLR}{Maximum-local-rate}
\newacronym{wbf}{WBF}{Wired Bias Function}
\newacronym{mib}{MIB}{Master Information Block}
\newacronym{sib}{SIB}{Secondary Information Block}
\newacronym{kpi}{KPI}{Key Performance Indicator}
\newacronym{ppp}{PPP}{Poisson Point Process}
\newacronym{gtp}{GTP}{GPRS Tunneling Protocol}
\newacronym{amf}{AMF}{Access and Mobility Management Function}
\newacronym{dash}{DASH}{Dynamic Adaptive Streaming over HTTP}
\newacronym{http}{HTTP}{Hypertext Transfer Protocol}
\newacronym{qos}{QoS}{Quality of Service}
\newacronym{udp}{UDP}{User Datagram Protocol}
\newacronym{cu}{CU}{Central Unit}
\newacronym{du}{DU}{Distributed Unit}
\newacronym{mt}{MT}{Mobile Termination}
\newacronym{sdap}{SDAP}{Service Data Adaptation Protocol}
\newacronym{tdm}{TDM}{Time Division Multiplexing}
\newacronym{fdm}{FDM}{Frequency Division Multiplexing}
\newacronym{sdm}{SDM}{Space Division Multiplexing}
\newacronym{dag}{DAG}{Directed Acyclic Graph}
\newacronym{st}{ST}{Spanning Tree}

\tikzstyle{startstop} = [rectangle, rounded corners, minimum width=2cm, minimum height=0.5cm,text centered, draw=black]
\tikzstyle{io} = [trapezium, trapezium left angle=70, trapezium right angle=110, minimum width=3cm, minimum height=1cm, text centered, draw=black]
\tikzstyle{process} = [rectangle, minimum width=2cm, minimum height=0.5cm, text centered, draw=black, alignb=center]
\tikzstyle{decision} = [ellipse, minimum width=2cm, minimum height=1cm, text centered, draw=black]
\tikzstyle{arrow} = [thick,<->,>=stealth]
\tikzstyle{line} = [thick,>=stealth]
\tikzstyle{darrow} = [thick,<->,>=stealth,dashed]
\tikzstyle{sarrow} = [thick,->,>=stealth]
\tikzstyle{larrow} = [line width=0.1mm,dashdotted,->,>=stealth]

\makeatletter
\def\grd@save@target#1{%
  \def\grd@target{#1}}
\def\grd@save@start#1{%
  \def\grd@start{#1}}
\tikzset{
  grid with coordinates/.style={
    to path={%
      \pgfextra{%
        \edef\grd@@target{(\tikztotarget)}%
        \tikz@scan@one@point\grd@save@target\grd@@target\relax
        \edef\grd@@start{(\tikztostart)}%
        \tikz@scan@one@point\grd@save@start\grd@@start\relax
        \draw[minor help lines] (\tikztostart) grid (\tikztotarget);
        \draw[major help lines] (\tikztostart) grid (\tikztotarget);
        \grd@start
        \pgfmathsetmacro{\grd@xa}{\the\pgf@x/1cm}
        \pgfmathsetmacro{\grd@ya}{\the\pgf@y/1cm}
        \grd@target
        \pgfmathsetmacro{\grd@xb}{\the\pgf@x/1cm}
        \pgfmathsetmacro{\grd@yb}{\the\pgf@y/1cm}
        \pgfmathsetmacro{\grd@xc}{\grd@xa + \pgfkeysvalueof{/tikz/grid with coordinates/major step x}}
        \pgfmathsetmacro{\grd@yc}{\grd@ya + \pgfkeysvalueof{/tikz/grid with coordinates/major step y}}
        \foreach \x in {\grd@xa,\grd@xc,...,\grd@xb}
        \node[anchor=north] at (\x,\grd@ya) {\pgfmathprintnumber{\x}};
        \foreach \y in {\grd@ya,\grd@yc,...,\grd@yb}
        \node[anchor=east] at (\grd@xa,\y) {\pgfmathprintnumber{\y}};
      }
    }
  },
  minor help lines/.style={
    help lines,
    gray,
    line cap =round,
    xstep=\pgfkeysvalueof{/tikz/grid with coordinates/minor step x},
    ystep=\pgfkeysvalueof{/tikz/grid with coordinates/minor step y}
  },
  major help lines/.style={
    help lines,
    line cap =round,
    line width=\pgfkeysvalueof{/tikz/grid with coordinates/major line width},
    xstep=\pgfkeysvalueof{/tikz/grid with coordinates/major step x},
    ystep=\pgfkeysvalueof{/tikz/grid with coordinates/major step y}
  },
  grid with coordinates/.cd,
  minor step x/.initial=.5,
  minor step y/.initial=.2,
  major step x/.initial=1,
  major step y/.initial=1,
  major line width/.initial=1pt,
}
\makeatother

\definecolor{desireRed}{RGB}{230,57,60}%
\definecolor{darkPurple}{RGB}{59,31,43}%
\definecolor{springGreen}{RGB}{37,223,145}%
\definecolor{queenBlue}{RGB}{69,123,157}%
\definecolor{spaceCadet}{RGB}{29,53,87}%

\makeglossaries

\begin{document}

\title{Integrated Access and Backhaul in 5G mmWave Networks: Potentials and Challenges}

\author{\IEEEauthorblockN{Michele Polese, \IEEEmembership{Student Member, IEEE}, Marco Giordani, \IEEEmembership{Student Member, IEEE}, Tommaso Zugno,\\Arnab Roy, \IEEEmembership{Member, IEEE}, Sanjay Goyal, \IEEEmembership{Member, IEEE}, Douglas Castor, \IEEEmembership{Senior Member, IEEE},\\Michele Zorzi, \IEEEmembership{Fellow, IEEE}}
\thanks{Michele Polese, Marco Giordani, Tommaso Zugno and Michele Zorzi are with the Department of Information Engineering (DEI), University of Padova, Italy, and Consorzio Futuro in Ricerca (CFR), Italy. Email:\{polesemi,giordani,zugnotom,zorzi\}@dei.unipd.it.

Arnab Roy, Sanjay Goyal and Douglas Castor are with InterDigital Communications, Inc., USA. Email:\{arnab.roy, sanjay.goyal, douglas.castor\}@interdigital.com.
}
}


\flushbottom
\setlength{\parskip}{0ex plus0.1ex}

\maketitle
\glsunset{nr}

\begin{abstract}
\gls{iab} is being investigated as a means to overcome deployment costs of ultra-dense 5G millimeter wave (mmWave) networks by realizing wireless backhaul links to relay the access traffic.
For the development of these systems, however, it is fundamental to validate the performance of IAB in realistic scenarios through end-to-end system level simulations.
In this paper, we shed light on the most recent standardization activities on IAB, and compare architectures with and without IAB in mmWave deployments.
While it is well understood that IAB networks reduce deployment costs by obviating the need to provide wired backhaul to each cellular base-station, in this paper we demonstrate the cell-edge throughput advantage offered by IAB using end-to-end system level simulations. We further highlight some research challenges associated with this architecture that will require further investigations.
\end{abstract}

\begin{IEEEkeywords}
5G, Millimeter Wave, Integrated Access and Backhaul, 3GPP, NR.
\end{IEEEkeywords}

\begin{picture}(0,0)(0,-380)
\put(0,0){
\put(0,0){\small This paper has been submitted to IEEE for publication. Copyright may be transferred without notice.}}
\end{picture}

\glsresetall

\section{Introduction}

Recently, the \gls{3gpp} has completed, as part of its Release 15, the standardization of a new \gls{rat}, i.e., 3GPP NR, that introduces novel designs and technologies to  comply with the ever more stringent performance requirements~\cite{38300} for \gls{5g} networks.
In addition to a flexible frame structure and a revised core network design, NR features carrier frequencies up to 52.6 GHz.
The large available spectrum at \gls{mmwave} frequencies, with bandwidths which are often much larger than in previous network generations, offers the potential of orders of magnitude higher transmission speeds than when operating in the congested bands below 6 GHz, and improves security and privacy because of the directional transmissions which are typically established~\cite{rangan2017potentials}.

However, operating in the \gls{mmwave} spectrum comes with its own set of challenges, severe path and penetration losses being one of them~\cite{rangan2017potentials}.
One promising approach to overcome such limitations is using high gain antennas to help close the link, thus introducing directionality in the communication, with electronic beamforming to support mobile users.
Network densification is also used to improve the performance by reducing inter-site distance to establish stronger access channels. An ultra-dense deployment, however, involves high capital and operational expenditures (capex and opex) for network operators~\cite{lopezperez2015towards}, because high capacity backhaul connections have to be provided to a larger number of cellular base stations than in networks operating at lower frequencies.

Network disaggregation (i.e., the separation of the layers of the protocol stack into different physical equipments)~\cite{makris2018cloud} and virtualization (i.e., the usage of software- and not hardware-based protocol stack implementations)~\cite{mijumbi2016network} can lower capex and opex by reducing the complexity of individual base stations.
Some researchers have also started investigating the feasibility of \gls{iab}, in which only a fraction of \gls{gnb} connect to traditional fiber-like infrastructures, while the others wirelessly relay the backhaul traffic, possibly through multiple hops and at \gls{mmwave} frequencies~\cite{dhillon2015wireless}.
The importance of the \gls{iab} framework as a cost-effective alternative to the wired backhaul has been recognized by the \gls{3gpp}. Indeed, it has recently completed a Study Item for 3GPP NR Release 16~\cite{38874}, which investigates architectures, radio protocols, and physical layer aspects for sharing radio resources between access and backhaul links. Although the \gls{3gpp} \gls{lte} and \gls{lte}-Advanced standards already provide specifications for base stations with wireless backhauling capabilities, the Study Item on \gls{iab} foresees a more advanced and flexible solution, which includes the support of multi-hop communications, dynamic multiplexing of the resources, and a plug-and-play design to reduce the deployment complexity. However, despite the consensus about  \gls{iab}'s ability to reduce costs, designing an efficient and high-performance IAB network is still an open research  challenge.

In this paper, we review 3GPP standardization activities on IAB and evaluate the performance of the IAB architecture in realistic deployments.
In particular, we compare network scenarios in which a percentage of gNBs (i.e., the \gls{iab}-nodes) use wireless backhaul connections to a few gNBs (i.e., the \gls{iab}-donors) with a wired connection to the core network against two baseline solutions, i.e., a network with only the \gls{iab}-donors, and one in which all the gNBs have a wired connection.
We also investigate how to efficiently forward the backhaul traffic from the wireless \gls{iab}-nodes to the core network and demonstrate the impact of topology setup strategies on the overall throughput and latency performance.
To do so, we conduct end-to-end simulations using ns-3, an open-source network simulator which has recently been extended to feature a detailed 3GPP-like protocol stack implementation of IAB at \glspl{mmwave}~\cite{polese2018end}.
Unlike traditional performance analyses, e.g.,~\cite{saha2018bandwidth,polese2018distributed,ometov2019nr}, which are focused on \gls{phy} or \gls{mac} layer layer protocols, we also investigate the impact of upper layers, thereby providing a more comprehensive network-level analysis. Moreover, we consider both traditional \gls{udp} services and more realistic applications including web browsing and \gls{dash} for high-quality video streaming.
Our results demonstrate that, while wired backhaul implementations deliver improved overall throughput in conditions of highly saturated traffic, the IAB configuration promotes fairness for the worst users by associating to relay nodes (\gls{iab}-nodes) the UEs which otherwise would have a poor connection to the wired donor.

\begin{figure*}[t]
	\centering
	\includegraphics[width=\textwidth]{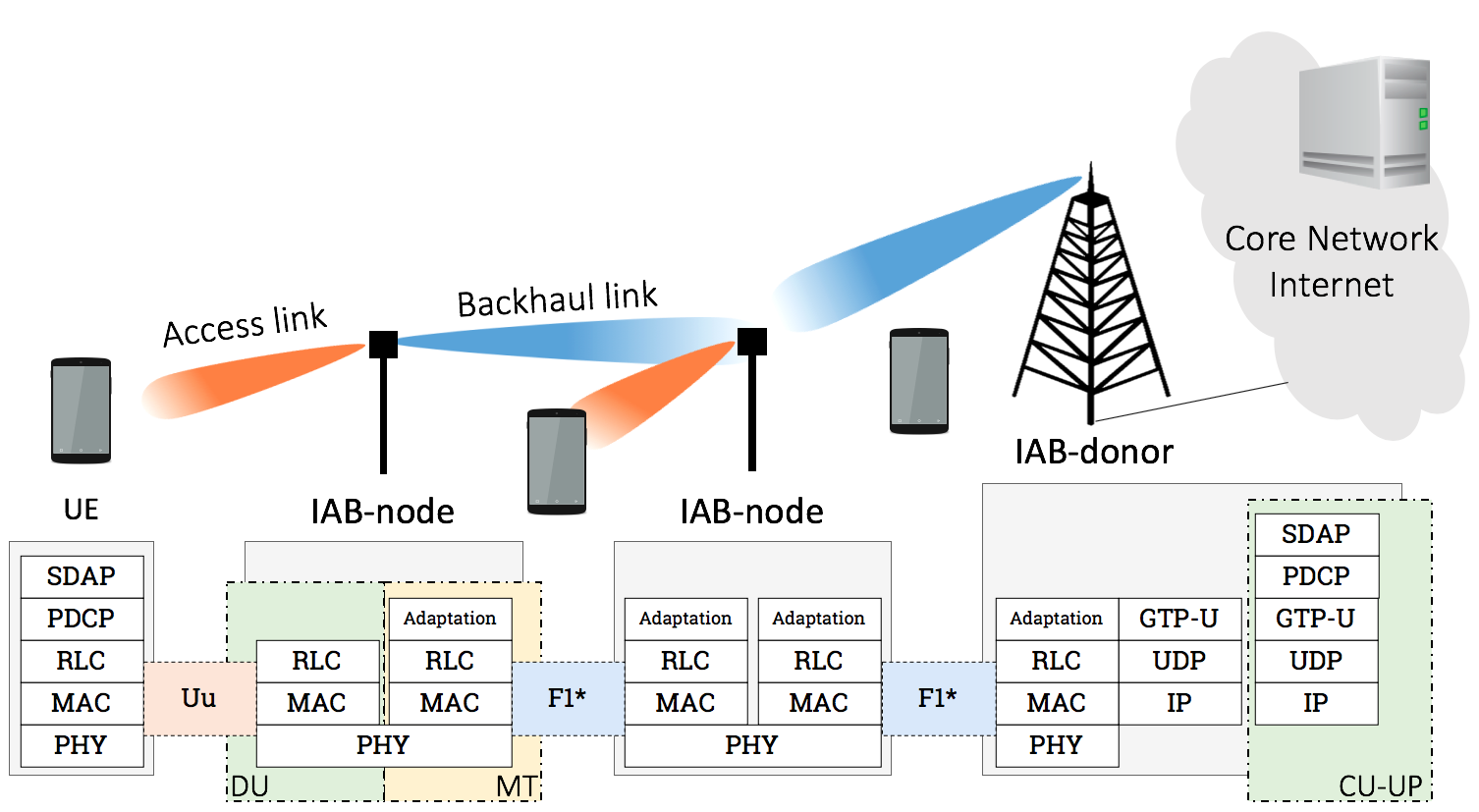}
	\caption{Protocol stack and basic architecture of an \gls{iab} network. The Uu interface represents the interface between the \gls{ue} and the \gls{du} in the \gls{iab}-node, while the F1$^*$ interface is used between the \gls{iab} \gls{du} and the upstream \gls{cu}.}
	\label{fig:iabnet}
\end{figure*}

Despite these encouraging features, real benefits of the IAB architecture and questions on network behavior under different traffic conditions have been largely unanswered so far. Accordingly, in this work we study the  performance of a typical network under different traffic considerations and
provide insights on the observed performance gains and shortcomings under different scenarios.

The remainder of the paper is organized as follows. In Sec.~\ref{sec:iab} we describe the \gls{3gpp} activities related to \gls{iab}. Then, in Sec.~\ref{sec:model}, we discuss the model and the results of the end-to-end performance evaluation, and identify \gls{iab} potentials and challenges in Sec.~\ref{sec:potentials}. Finally, we conclude the paper in Sec.~\ref{sec:concl}.

\section{Integrated Access and Backhaul in 3GPP NR}
\label{sec:iab}

The \gls{3gpp} recently finalized the Study Item on \gls{iab} \cite{38874}, whose main objective was to assess the feasibility of integrated access and wireless backhaul over NR (i.e., the 5G radio interface), and to propose potential solutions to ensure efficient backhauling operations. This Study Item led to a Work Item, and is expected to be integrated in future releases of the \gls{3gpp} specifications.

The Study Item considered fixed wireless relays with both in-band (i.e., the access and the backhaul traffic are multiplexed over the same frequency band) and out-of-band backhauling capabilities (i.e., the access and the backhaul traffic use separate frequency bands), with a focus on the former, which is more challenging in terms of network design and management but maximizes the spectrum utilization. According to~\cite{38874}, \gls{iab} operations are spectrum agnostic, thus the relays can be deployed either in the above-6 GHz or sub-6 GHz spectrum, and can operate both in \gls{sa} (connected to the 5G core network) or \gls{nsa} modes (connected to the 4G \gls{epc}). The possible topologies for an \gls{iab} network are (i) a \gls{st}, in which each \gls{iab}-node is connected to a single parent, or (ii) a \gls{dag}, in which each \gls{iab}-node may be connected to multiple upstream nodes.

In the following sections, we will review the main innovations introduced in~\cite{38874} for the network architecture, the procedures for network management, and the resource multiplexing through scheduling.

\glsreset{ue}

\subsection{Architecture}
As shown in Fig.~\ref{fig:iabnet}, the logical architecture of an \gls{iab} network is composed of multiple \gls{iab}-nodes, which have wireless backhauling capabilities and can serve \glspl{ue} as well as other \gls{iab}-nodes, and \gls{iab}-donors, which have fiber connectivity towards the core network and can serve \glspl{ue} and \gls{iab}-nodes. 

The Study Item initially proposed five different configuration options for the architecture, with different levels of decentralization of the network functionalities and different solutions to enable backhauling.
The final version, selected for future standardization, was preferred because it had limited impact on the core network specifications, had lower relay complexity and processing requirements, and had more limited signaling overhead.

\glsreset{du}

According to the chosen architecture, each \gls{iab}-node hosts two NR functions: (i) a \gls{mt}, used to maintain the wireless backhaul connection towards an upstream \gls{iab}-node or \gls{iab}-donor, and (ii) a \gls{du}, to provide access connection to the \glspl{ue} or the downstream \glspl{mt} of other \gls{iab}-nodes. 
The \gls{du} connects to a \gls{cu} hosted by the \gls{iab}-donor by means of the NR F1$^*$ interface running over the wireless backhaul link. Therefore, in the access of \gls{iab}-nodes and donors there is a coexistence of two interfaces, i.e., the Uu interface (between the \glspl{ue} and the \gls{du} of the gNBs) and the aforementioned F1$^*$ interface.

With this choice it is possible to exploit the functional split of the radio protocol stack: the \gls{cu} at the \gls{iab}-donor holds all the control and upper layer functionalities, while the lower layer operations are delegated to the \glspl{du} located at the \gls{iab}-nodes. The split happens at the \gls{rlc} layer, therefore \gls{rrc}, \gls{sdap} and \gls{pdcp} layers reside in the \gls{cu}, while \gls{rlc}, \gls{mac} and \gls{phy} are hosted by the \glspl{du}. An additional adaptation layer is added on top of \gls{rlc}, which routes the data across the \gls{iab} network topology, hence enabling the end-to-end connection between \glspl{du} and the \gls{cu}.

\subsection{Network Procedures and Topology Management}
\label{ssec:net_proc}

An important element to be considered in an \gls{iab} deployment is the establishment and management of the network topology. This is because the end-to-end performance of the overall network strongly depends on the number of hops between the donor and the end relay, on how many relays the donor needs to support, and strategies adopted for procedures such as network formation, route selection and resource allocation.
To ensure efficient \gls{iab} operations, it is necessary to optimize the performance of various network procedures involving topology and resource management.

The topology establishment is performed during the \gls{iab}-node setup, and is a critical step.
When an \gls{iab}-node becomes active, it first selects the upstream node to attach to. To accomplish this, the \gls{mt} performs the same initial access procedure as a \gls{ue}, i.e., it makes use of the synchronization signals transmitted by the available cells (formally called synchronization signal block (SSB) in NR) to estimate the channel and select the parent. Moreover, although not currently supported by the specifications, we argue that it would be beneficial if the \gls{mt} could retrieve additional information (e.g., the number of hops to reach the donor, the cell load, etc.), and then select the cell to attach to, based on more advanced path selection metrics~\cite{giordani2018tutorial} than just the \gls{rss}, as will be discussed in Sec.~\ref{sec:model}. Then, the \gls{iab}-node configures its \gls{du}, establishes the F1$^*$ connection towards the \gls{cu} in the remote \gls{iab}-donor, and is ready to provide services to \glspl{ue} and other \gls{iab}-nodes. During this initial phase, the \gls{iab}-node may transmit information to the \gls{iab}-donor about its topological location within the \gls{iab} network.

The topology management function then dynamically adapts the \gls{iab} topology in order to maintain service continuity (e.g., when a backhaul link is degraded or lost), or for load balancing purposes (e.g., to avoid congestion).
In addition to the information provided during the initial setup procedure, the \gls{iab}-nodes may also transmit periodic information about traffic load and backhaul link quality.
This allows the \gls{cu} to be aware of the overall \gls{iab} topology, find the optimal configuration, and adapt it by changing network connectivity (i.e., the associations between the \gls{iab}-nodes) accordingly.

In case the \gls{iab}-nodes support a \gls{dag} topology with multi-connectivity towards multiple upstream nodes, it is also possible to provide greater redundancy and load balancing. In this case, the addition/removal of redundant routes is managed by the \gls{cu} based on the propagation conditions and traffic load of each wireless backhaul link.

\begin{figure*}
	\centering
	\includegraphics[width=0.99\textwidth]{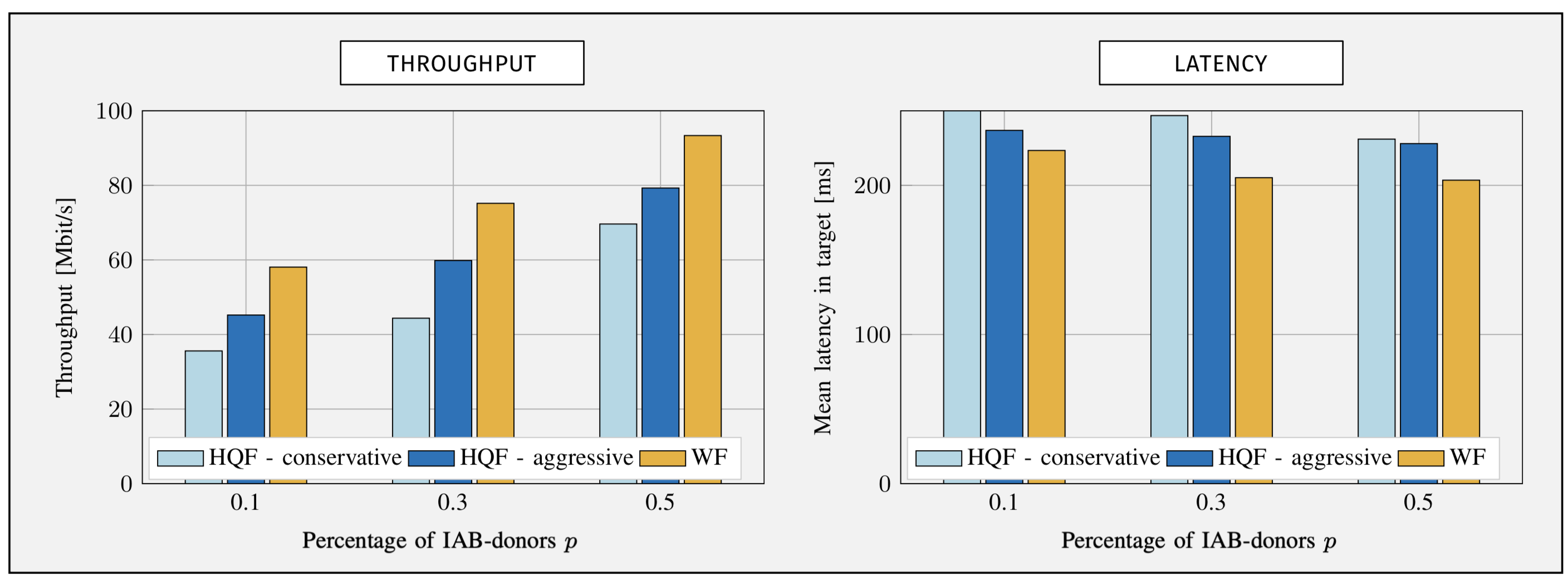}
	\caption{Throughput (left) and latency (right) comparison of \gls{iab} path selection policies varying the percentage of \gls{iab}-donors $p$ for a density of 45 gNB/km$^2$ and a constant bitrate traffic.}
	\label{fig:policies}
\end{figure*}

\subsection{Scheduling and Resource Multiplexing}
For in-band \gls{iab} operations, the need to multiplex both the access and the backhaul traffic within the same frequency band forces half-duplex operations. This constraint has been acknowledged in the 3GPP Study Item report~\cite{38874}, although full-duplex solutions are not excluded. Therefore, the radio resources must be orthogonally partitioned between the access and the backhaul, either in time (\gls{tdm}, which is the preferred solution in~\cite{38874}), frequency (\gls{fdm}), or space (\gls{sdm}), using a centralized or decentralized scheduling coordination mechanism across the \gls{iab}-nodes and the \gls{iab}-donor.

Despite the limitations imposed by the half-duplex constraint, the \gls{iab} network is required to address the access traffic requirements of all the users. For this reason, the available resources should  be  allocated fairly, taking into account channel measurements and topology-related information possibly exchanged between the \gls{iab}-nodes. Furthermore, both hop-by-hop and end-to-end flow control mechanisms should be provided to mitigate the risk of congestion on intermediate hops, which might arise in case of poor propagation conditions.

\section{End-to-end Evaluation of IAB}
\label{sec:model}

We have evaluated the end-to-end performance of an \gls{iab} mmWave network using the simulator described in~\cite{polese2018end}, which implements a full-stack model of the cellular network, with the 3GPP channel model for mmWave frequencies and beamforming. Moreover, thanks to the integration with ns-3, it is possible to study end-to-end scenarios with the TCP/IP stack~\cite{henderson2008network} and realistic applications, such as the 3GPP \gls{http} model. In the scenario we investigated, the base stations are deployed following a \gls{ppp} with density $\lambda$ BS/km$^2$, and a fraction $0 \le p \le 1$ of the $N$ base stations have wired backhaul connections (i.e., the \gls{iab}-donors), while the others (i.e., the \gls{iab}-nodes) are wirelessly connected to the \gls{iab}-donors, perhaps over multiple hops. The network implements in-band backhaul, at 28 GHz, with \gls{tdm} of the radio resources among the access and the backhaul links. We consider uniform rectangular antenna arrays in the base stations and \glspl{ue}, with 64 and 16 elements, respectively, and the beamforming model described in~\cite{mezzavilla2018end}. The base stations use the backhaul-aware round robin scheduler presented in~\cite{polese2018end}. 
The \glspl{ue} are also deployed with a \gls{ppp} with density $\lambda_u = 10 \lambda$ \gls{ue}/km$^2$, although we only evaluate the performance of the subset of users connected to a target base station, which is either the first gNB deployed in a baseline scenario in which all nodes have a wired connection to the core network, or the first \gls{iab}-node that performs the initial access in an \gls{iab} scenario.

\begin{figure*}
	\centering
	\begin{subfigure}[t]{0.49\textwidth}
		\centering
	  	\includegraphics[width=0.99\textwidth]{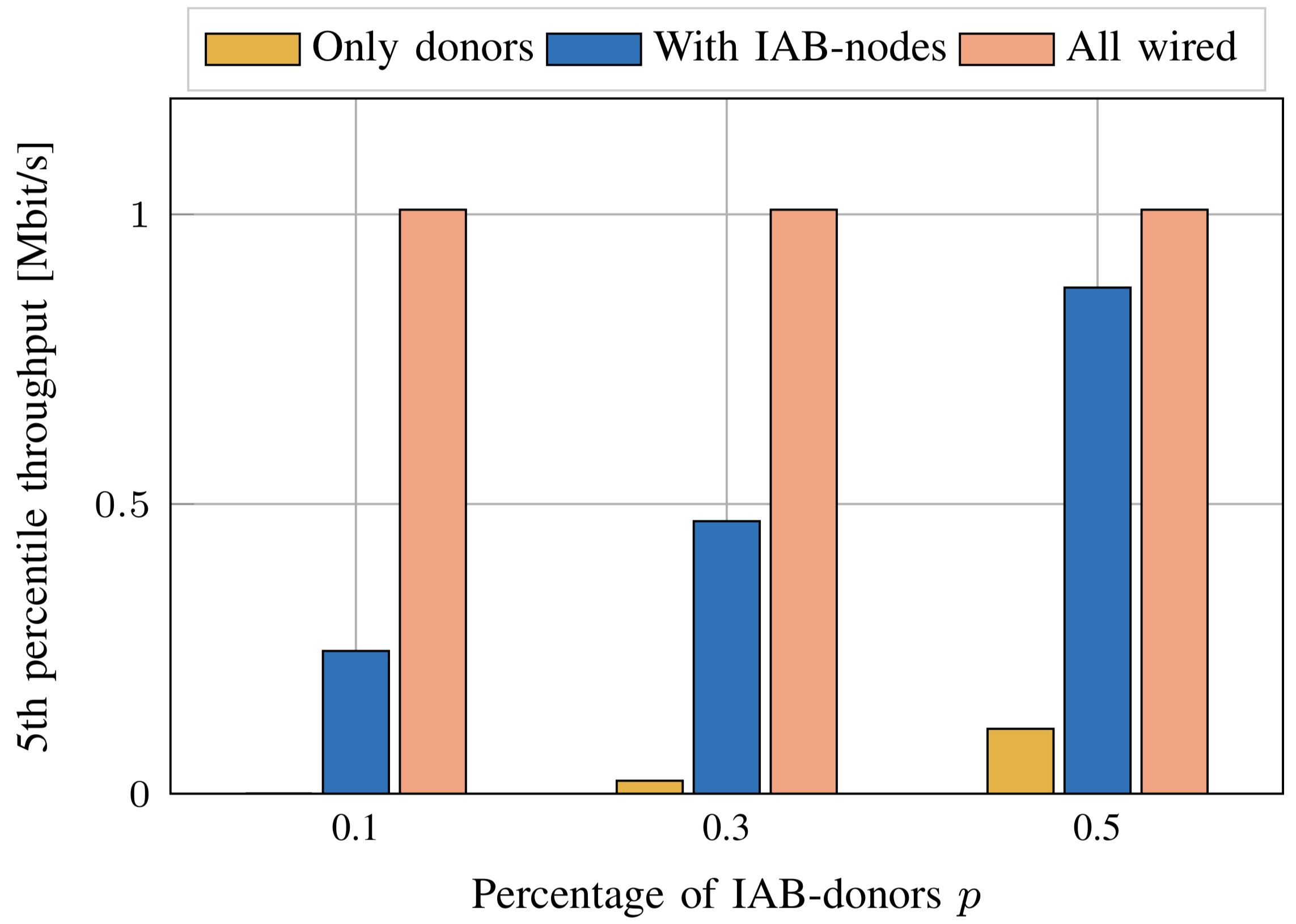}
	  	\caption{Fifth percentile throughput.}
  		\label{fig:thIab}
	\end{subfigure}\hfill
	\begin{subfigure}[t]{0.49\textwidth}
		\centering
	  	\includegraphics[width=0.99\textwidth]{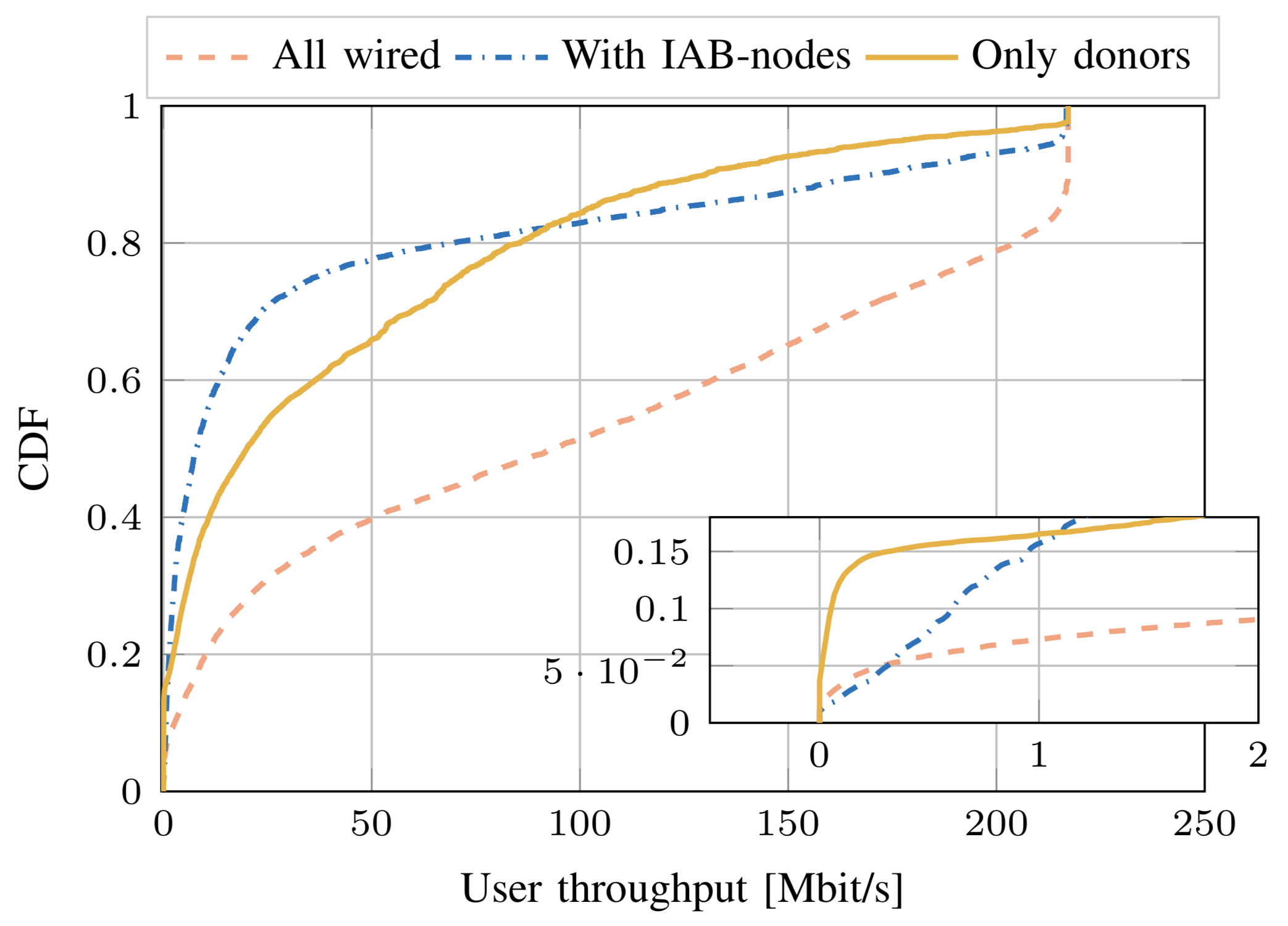}
	  	\caption{Throughput \gls{cdf} for $p=0.3$.}
  		\label{fig:cdf}
	\end{subfigure}
	\caption{Fifth percentile and \gls{cdf} of the throughput for the users of a target \gls{iab}-node, varying the percentage of \gls{iab}-donors $p$ and the deployment strategies, for a density of 45 gNB/km$^2$.}
	\label{fig:udp}
\end{figure*}

\smallskip
\textbf{Backhaul path selection policies.} The first set of results, reported in Fig.~\ref{fig:policies}, sheds light on the impact of different backhaul path selection policies in an \gls{iab} setup.
As introduced in Sec.~\ref{ssec:net_proc},  path selection refers to the procedure by which \gls{iab}-nodes find the path towards an \gls{iab}-donor, possibly through multiple hops.
In our previous work~\cite{polese2018distributed}, we investigated two different policies to forward the backhaul traffic: (i)  a \emph{\gls{hqf}} approach which selects, as a parent, the gNB with the highest quality, i.e., the highest \gls{snr}, and (ii) a \emph{\gls{wf}} approach which selects a direct link to the \gls{iab}-donor with the best signal, even if an IAB-node with better channel quality is available, provided that some minimum channel quality criterion is satisfied.
The first approach facilitates a best-quality wireless backhaul connection in the first hop but, in turn, may increase the number of hops required to forward the traffic to an \gls{iab}-donor.
The second approach,  while minimizing the number of end-to-end hops, may choose backhaul links with poorer channel quality.
The \gls{hqf} policy may also leverage a function that biases the link selection towards gNBs with wired backhaul to decrease the number of hops to the core network.
The bias computed by the function is not fixed, but depends on the number of hops from the IAB-node to the candidate parent it is trying to connect to~\cite{polese2018distributed}.
Moreover, both conservative and aggressive bias functions can be designed (aggressive \gls{hqf} policies will progressively operate like  \gls{wf} policies).
Fig.~\ref{fig:policies} demonstrates that the \gls{wf} approach should be preferred since it  offers lower end-to-end latency and higher total throughput compared to the other investigated policies. The results show that minimizing the number of hops required to connect to an \gls{iab}-donor improves throughput and reduces latency by reducing relaying overhead and congestion at intermediate \gls{iab}-nodes.

\begin{figure*}
	\centering
	\begin{subfigure}[t]{0.49\textwidth}
		\centering
	  	\includegraphics[width=0.99\textwidth]{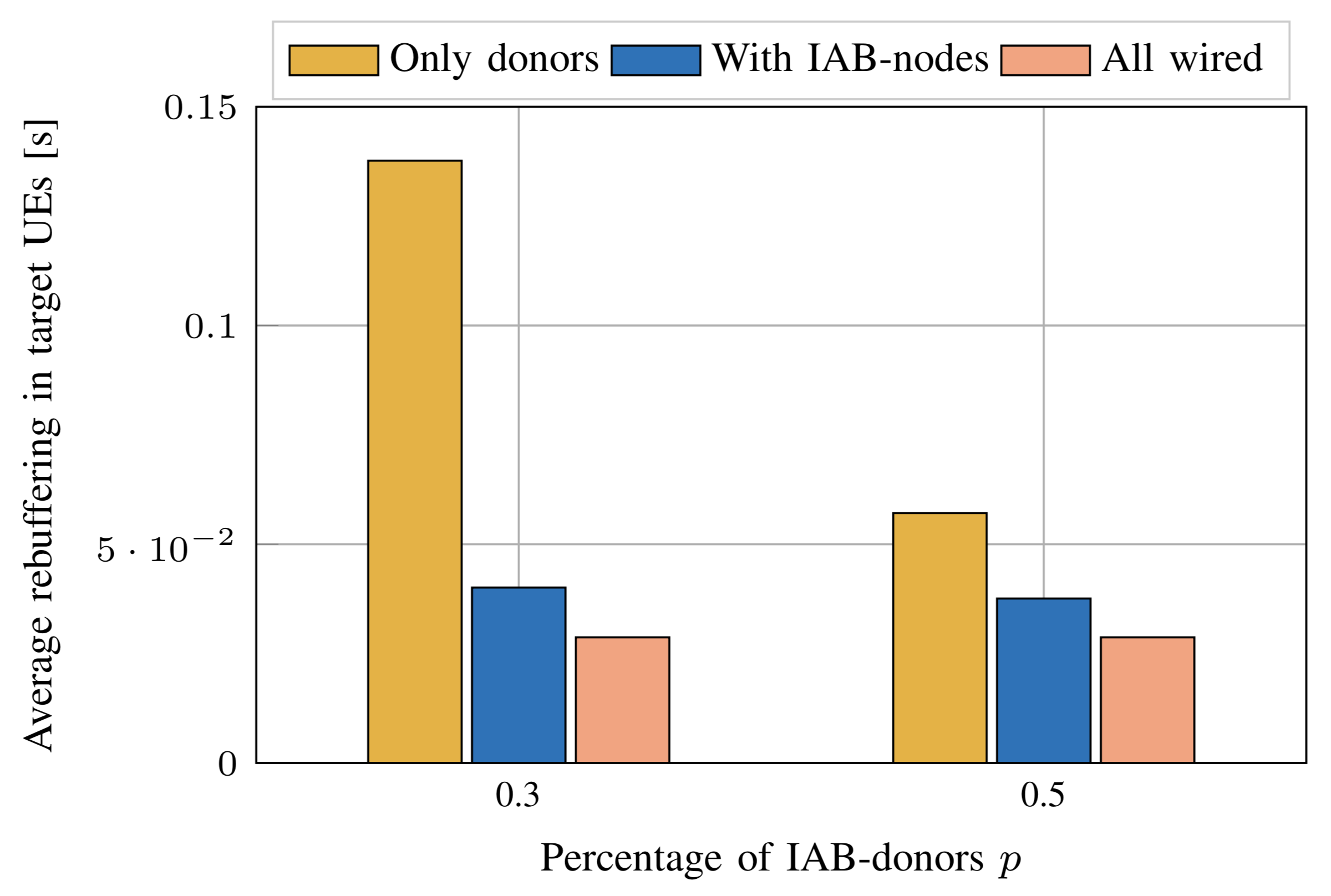}
	  	\caption{Average rebuffering for DASH clients.}
  		\label{fig:dash}
	\end{subfigure}\hfill
	\begin{subfigure}[t]{0.49\textwidth}
		\centering
	  	\includegraphics[width=0.99\textwidth]{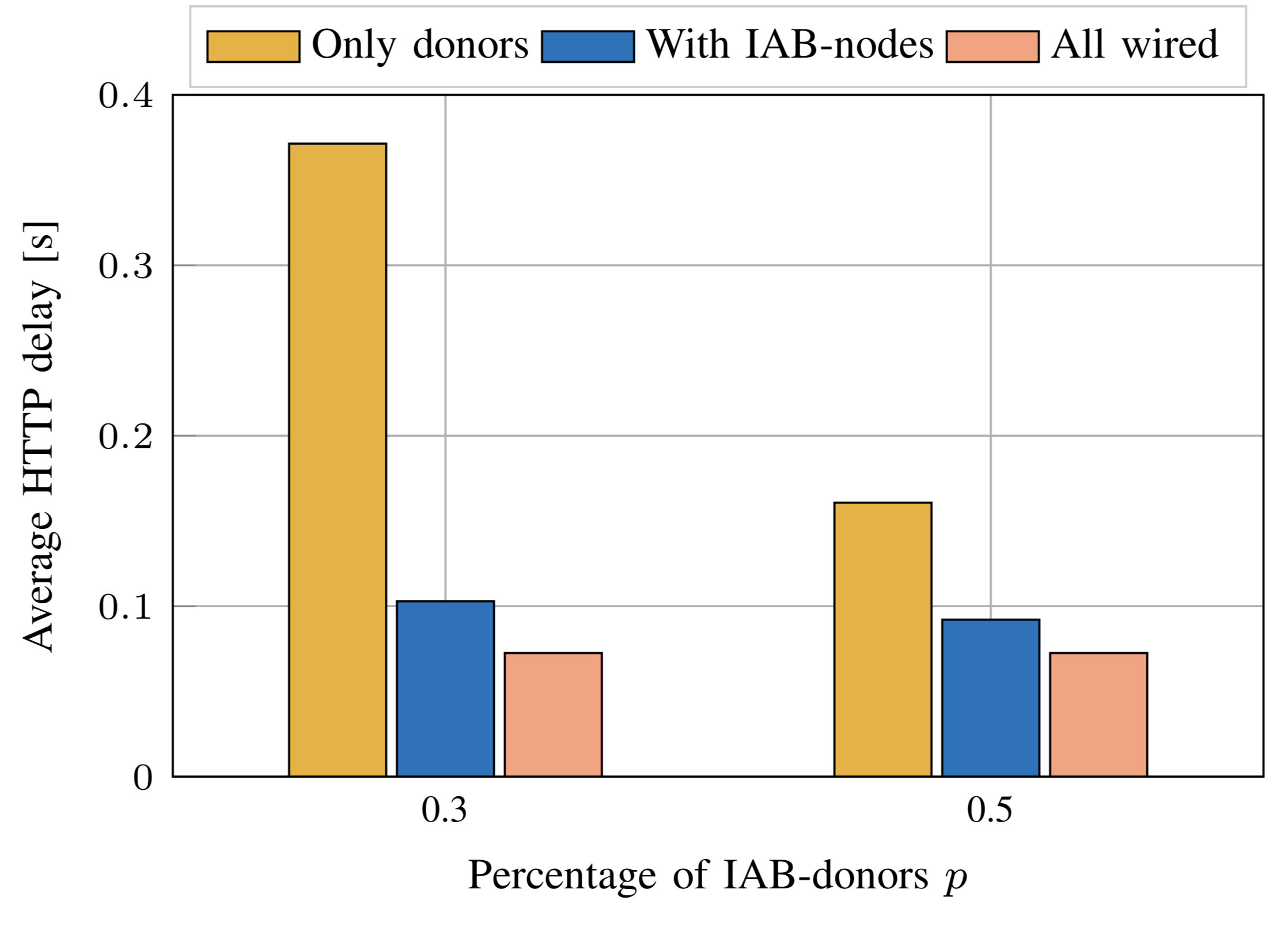}
	  	\caption{Average delay to retrieve an HTTP web page.}
  		\label{fig:http}
	\end{subfigure}
	\caption{Performance for users in a target \gls{iab}-node, with different applications, for a density of 30 gNB/km$^2$.}
	\label{fig:apps}
\end{figure*}

\smallskip
\textbf{IAB deployment scenarios.} We also tested three different deployment scenarios. The best case is when all the $N$ base stations in the network are equipped with a wired connection to the core network (i.e., the \textit{all wired} scenario). This represents the most expensive solution, in terms of density of fiber drops, but permits the whole bandwidth to be used for access traffic. With the \textit{\gls{iab}-nodes} option, $pN$ base stations are \gls{iab}-donors, i.e., have a wired connection and $(1-p)N$ have wireless backhaul. Finally, the baseline is the one that 3GPP considers for comparisons with \gls{iab} solutions, described in~\cite{38874}, i.e., a deployment with only $pN$ wired base stations and no \gls{iab}-nodes (the \textit{only donors} configuration).

\textit{UDP user traffic.} In Fig.~\ref{fig:udp}, we consider an \gls{iab} network where each user downloads content from a remote server with a constant bitrate of 220 Mbps, using UDP as the transport protocol, thus introducing a full buffer source traffic model. The flow of each end-to-end connection does not self-regulate to the actual network conditions, thus congestion arises. This experiment aims to test the performance of an \gls{iab} setup in a saturation regime, where the access and backhaul links are constantly used. As expected, the best performance is provided by the all wired configuration, given that it provides the same access point density as the \gls{iab} setup, but avoids the multiplexing of resources between access and backhaul. On the other hand, it is possible to identify two advantages and one drawback of the \gls{iab} configuration with respect to the only donors one. A higher throughput for the worst users is achieved when using \gls{iab}-nodes, as shown by the fifth percentile throughput plot in Fig.~\ref{fig:thIab}. In particular, for $p=0.5$ (i.e., when the number of relays is equal to the number of \gls{iab}-donors), \gls{iab} has only 13\% less fifth percentile throughput than the all wired configuration. Moreover, the usage of \gls{iab}-nodes likely offloads the worst users from the \gls{iab}-donors, and this frees up resources for users with the best \gls{iab}-donor channel quality, thereby enabling a higher throughput, as illustrated in Fig.~\ref{fig:cdf}. The \gls{iab} solution, however, requires multiplexing of the wireless resources between access and backhaul. In a scenario where the links are always saturated, this results in a worse performance for the average users connected to the relays, which are throttled on the backhaul links by the round robin scheduler at the donors and have a smaller throughput than with the only donors setup.

\smallskip
\textit{DASH, HTTP user traffic.}
The next set of results considers a more common use case, in which the users either stream video using \gls{dash}~\cite{stockhammer2011dash} or access web pages using \gls{http} from a remote server. This kind of source traffic is asynchronous and bursty, and, in the \gls{dash} case, the flow adapts itself to the varying capacity offered by the network, after some delays due to the signaling and convergence of the algorithm. 
Therefore, the network is not as stressed as in the previous experiment, and in this case  the advantage of \gls{iab} is more visible. Indeed, thanks to the better channel seen on average by the user due to more numerous nodes compared to the only donor case, and thanks to the asynchronous and independent nature of the traffic at each user, which provides greater multiplexing gains, the performance of the \gls{iab} network is not far from that of the network with all wired access points. In particular, Fig.~\ref{fig:dash} reports the average duration of a rebuffering event for a \gls{dash} stream, for all the users in a target base station. The rebuffering happens when the \gls{dash} framework does not adapt fast enough to the network conditions, or if the network capacity is not sufficient to sustain even the minimum video quality available in the \gls{dash} remote server. As can be seen, the only donors setup has the worst performance, with a 5 and 2 times higher rebuffering than the all wired configuration, for $p=0.3$ and 0.5, respectively. The \gls{iab} deployment, instead, degrades the performance of the all wired only by 1.4 and 1.3 times, for $p=0.3$ and 0.5, respectively. Likewise, Fig.~\ref{fig:http} shows the average time it takes to completely download a web page, from the first client \gls{http} request to the reception of the last object, and, as can be seen, the trend is similar to that of the \gls{dash} rebuffering. Finally, for this kind of traffic, the improvement introduced by the densification of \gls{iab}-donors (i.e., by increasing $p$ from 0.3 to 0.5) is less marked than with the constant bitrate traffic shown in Fig.~\ref{fig:udp}.

\section{Potentials and Challenges of \gls{iab}}
\label{sec:potentials}

As highlighted by the results presented in Sec.~\ref{sec:model}, \gls{iab} networks present both benefits and limitations with respect to deployments where the radio resources are not multiplexed between the access and the backhaul. 
First, the \gls{iab} solution may present lower deployment costs and complexity with respect to the all wired setup, but, at the same time, splitting the available resources between access and backhaul traffic makes the overall network performance  worse than in the all wired case under heavily loaded network scenarios. However, for bursty traffic the performance of the \gls{iab} solution approaches that of the all wired case. This shows that when evaluating the performance of \gls{iab} networks it is important to consider the specific use case and end-to-end applications that run on top of the network. Moreover, the results suggest that the main advantages of an \gls{iab} deployment, when compared to the only donors setup, come from an improvement in channel quality for cell edge users, on average, which consequently improves the area spectral efficiency.

On the other hand, the deployment of an \gls{iab} network presents challenges related to the design and interactions at different layers of the protocol stack. An important issue is related to the enforcement of \gls{qos} guarantees in single and multi hop scenarios, so that mixed IAB traffic flows for end-to-end applications can safely coexist. Additionally, the resources in the \gls{iab} network are limited and shared between the access and the backhaul. Therefore, the admission and bearer configuration should take this into consideration, in order to avoid overbooking the available resources and introducing congestion in the network. As shown in Fig.~\ref{fig:cdf}, this may indeed worsen the experience of the average users. Similarly, during the setup phase, in which the \gls{iab}-nodes join the network by performing initial access to their \gls{iab} parents, it is important to consider the attachment strategies to avoid overloading some \gls{iab}-donors, or excessively increasing the number of hops.
Even though we demonstrated in Fig.~\ref{fig:policies} that reducing the number of relay operations is always beneficial  in terms of both end-to-end latency and throughput, how to design path selection strategies which are robust to network topology changes and end terminals' mobility is still an open research challenge which deserves further investigation.

Most of these system-level challenges are strictly related to the design of ad hoc scheduling procedures at the \gls{mac} layer, able to efficiently  split the resources between the access and the backhaul and provide interference management. Another important challenge is related to cross-layer effects emerging from retransmissions at multiple layers, and the configuration of \gls{rlc} and transport layer timers may need to account for the additional delays related to the retransmissions over multiple hops and the reordering of packets at the receiver. At the physical layer, it will be interesting to evaluate the gain of the spatial multiplexing of the access and the backhaul, by using digital or hybrid beamforming, which could avoid the time or frequency multiplexing that are needed when using single-beam analog beamforming.

Overall, these challenges represent promising research directions to enable self-configuring, easy-to-deploy and high-performing \gls{iab} networks, which could represent a cost-effective solution for an initial ultra-dense NR deployment at mmWave frequencies.

\section{Conclusions and Future Work}
\label{sec:concl}
High-density deployments of 5G cells operating at \glspl{mmwave} call for innovative solutions to reduce capital and operating costs without degrading the end-to-end network performance. In this context, \gls{iab} has been investigated as an approach to relay access traffic to the core network wirelessly, thereby removing the need for all base stations to be equipped with fiber backhaul.
In this work we have reviewed the characteristics of IAB capabilities that are currently being standardized in 3GPP NR Release 16 and  evaluated the performance of \gls{iab} networks for different applications and traffic types such as Internet browsing (i.e., HTTP) and video streaming (i.e., DASH). We showed that  IAB  represents a viable solution to  efficiently relay  cell-edge  traffic, although the benefits decrease for more congested networks. We have also highlighted the limitations of the IAB paradigm and provided guidelines on how to overcome them.

IAB standardization is, however, still an on-going process.
As part of our future work, we will validate wireless backhaul solutions considering recently proposed 3GPP scenarios and investigate the impact of mobility and network reconfiguration on the network performance.

\bibliographystyle{IEEEtran}
\bibliography{bibl.bib}

\begin{IEEEbiographynophoto}{Michele Polese}
[S'17] received his B.Sc. (2014) and M.Sc. (2016) in Telecommunication Engineering from the University of Padova, Italy. Since October 2016 he has been a Ph.D. student at the University of Padova. He visited New York University (NYU), AT\&T Labs, and Northeastern University. His research focuses on protocols and architectures for 5G mmWave networks.
\end{IEEEbiographynophoto}
\begin{IEEEbiographynophoto}{Marco Giordani}
[S’17] received his B.Sc. (2013) and M.Sc. (2015) in Telecommunication Engineering from the University of Padova, Italy.
Since October 2016, he has been a Ph.D. student at the University of Padova.
He visited  New York University (NYU), USA, and TOYOTA Infotechnology Center, Inc., USA.
In 2018 he received the “Daniel E. Noble Fellowship Award” from the IEEE Vehicular Technology Society. His research  focuses on protocol design for 5G cellular and vehicular networks at mmWaves.
\end{IEEEbiographynophoto}
\begin{IEEEbiographynophoto}{Tommaso Zugno}
received his B.Sc. (2015) and M.Sc. (2018) in Telecommunication Engineering from the University of Padova, Italy. From May to October 2018 he was a Postgraduate researcher with the Department of  Information Engineering, University of Padova, under the supervision of Prof. Michele Zorzi. Since October 2018 he has been a Ph.D. student at the same university. His research focuses on protocols and architectures for 5G mmWave networks.
\end{IEEEbiographynophoto}
\begin{IEEEbiographynophoto}{Arnab Roy}
received his B.E. in Electronics Engineering from University of Mumbai, India, and his M.S. and Ph.D. in Electrical Engineering from Penn State University. He is currently with InterDigital, where he is working on technology development for next generation cellular and WLAN systems. Before joining InterDigital he held research and product development roles in the wireless communications industry. His current interests include communication technologies related to millimeterwave frequencies, unlicensed spectrum and automobiles.
\end{IEEEbiographynophoto}
\begin{IEEEbiographynophoto}{Sanjay Goyal}
received his B.Tech. (2009) in communication and computer engineering  from  the  LNM  Institute  of  Information Technology, India, and the Ph.D. (2016) and M.S. (2012) degrees in electrical engineering from  NYU  Tandon  School  of  Engineering,  New  York. Currently,  he  is working with  InterDigital  Communications, involved in technology development for next generation communication systems. He is a co-winner of a Best Paper Award from IEEE ICC 2016. His current research interests include NR in unlicensed spectrum, millimeter wave communications, full duplex cellular systems.
\end{IEEEbiographynophoto}
\begin{IEEEbiographynophoto}{Douglas Castor} [SM'19]
received his B.S. in Electrical Engineering from Pennsylvania State University in 1992 and his M.S. in Electrical Engineering from University of Pennsylvania in 1995. He is currently a Senior Director of R\&D at InterDigital, where he leads the incubation and development of emerging wireless and sensor technologies. Since joining InterDigital in 2000, he has supported roles in both product development and research innovations for 3G, 4G, 5G cellular and other wireless technologies.
\end{IEEEbiographynophoto}
\begin{IEEEbiographynophoto}{Michele Zorzi}
[F'07] is a Full Professor in the Information Engineering Department of the University of Padova. His present research interests focus on various aspects of wireless communications. He was Editor-in-Chief of IEEE Wireless Communications from 2003 to 2005, IEEE Transactions on Communications from 2008 to 2011, and IEEE Transactions on Cognitive Communications and Networking from 2014 to 2018. He served as a Member-at-Large of the ComSoc Board of Governors from 2009 to 2011, and as Director of Education and Training from 2014 to 2015.
\end{IEEEbiographynophoto}

\end{document}